# Large non-reciprocal charge transport in $Pt_2MnGe$ up to room temperature

K. K. Meng[1*], K. Wang[1], N. N. Zhang[2], Z. G. Fu[2], J. K. Chen[1], Y. Wu[1], X. G. Xu[1], J. Miao[1] and Y. Jiang[1,**]

[1]*School of Materials Science and Engineering, University of Science and Technology Beijing, Beijing 100083, China*

[2]*Institute of Applied Physics and Computational Mathematics, Beijing, China*

**Non-reciprocal charge transport that is strongly associated with the structural or magnetic chirality of the quantum materials system is one of the most exotic properties of condensed matter physics. Here, using magnetic alloys film $Pt_2MnGe$, we have realized the large non-reciprocal charge transport up to room temperature, which roots in the organic combination of chirality dependent carrier scattering and special magnetic configurations. In this framework, the conduction electrons are scattered asymmetrically by the emerging non-zero vector spin chirality under in-plane magnetic field, resulting in robust non-reciprocal response. More astonishingly, the vector spin chirality in $Pt_2MnGe$ film can be reversed by a spin-polarized current through spin Hall effect in a junction with Pt layer. Our work resolves the general limitation of non-reciprocal charge transport to cryogenic temperatures, and paves the way for extending its applications in the emerging field of chiral spintronics.**

[*]e-mail: kkmeng@ustb.edu.cn;
[**]e-mail: yjiang@ustb.edu.cn.

Chirality, which is characterized by a reflection asymmetry in terms of left hand being the mirror opposite of right hand, plays an essential role in modern condensed matter physics [1-3]. The interplay between structural chirality and conduction electrons in the conductors with broken spatial inversion symmetry could give rise to a non-reciprocal response [4, 5]. The roots of this non-reciprocal charge transport can be traced to the discovery of rectification effect in a metal/semiconductor heterostructures. It has stimulated experimental and theoretical interest for the wide use of electronic components, such as rectifiers and alternating-direct-current converters. Such a non-reciprocal response in the quantum material that the time-reversal symmetry is further broken by applying a magnetic field or spontaneous magnetization is often referred to as electrical magnetochiral anisotropy (EMCA), providing an important tool for investigating the magnetic chirality. The EMCA manifest itself as the appearance of bilinear magneto-electric resistance depending on the inner product of current and magnetic field, which therefore deviates strongly from the Onsager reciprocal theorem [6-13]. Consequently, the longitudinal resistance is given by [4, 5]:

$$R(\boldsymbol{I}, \boldsymbol{B}) = R_0(1 + \beta \boldsymbol{B}^2 + \gamma \boldsymbol{I} \cdot \boldsymbol{B}) \quad (1)$$

where $R_0$, $\boldsymbol{I}$, and $\boldsymbol{B}$ represent the resistance at zero magnetic field, the electric current, and magnetic field, respectively. The parameter $\beta$ describes the normal magnetoresistance that is allowed in all conductors, and $\gamma$ is the EMCA coefficient and its sign is associated with chirality. While the non-reciprocal charge transport has now been observed in a rich variety of quantum materials, it is still in its infancy. Several major issues such as the general limitation to cryogenic temperatures have to be resolved to realize its application in functional electronics devices. Furthermore, extending the non-reciprocal response to more regimes in modern condensed matter physics such as spintronics will explore a fertile ground for a great diversity of technological applications. Of particular interest here is the link between chirality and magnetic configurations. Recently, the non-collinear magnetic materials with non-zero spin chirality have brought about rich electronic and magnetic properties, such as large anomalous Hall effect (AHE) driven by Berry curvature in real or momentum space [14, 15], skyrmion lattices [16, 17], manipulation of the magnetic order parameter by spin-orbit torque (SOT) and rich spin Hall properties [18, 19].

In this work, we have realized the robust non-reciprocal charge transport in Pt$_2$MnGe (PMG) single layer up to room temperature, which roots in the organic combination of chirality dependent carrier scattering and special magnetic configurations. The magnetic properties of PMG have been investigated through magnetic curves, x-ray magnetic circular dichroism (XMCD) and x-ray magnetic linear dichroism (XMLD) measurements, which have verified a special magnetic configuration of PMG film up to room temperature. Both AC current harmonic resistance and pulsed current resistance measurements have confirmed the non-reciprocal charge transport in PMG. Large EMCA coefficient in the whole temperature from 5 K to 400 K indicates the strong correlation between conduction electrons and magnetic-field driven vector spin chirality in PMG. Furthermore, this kind of chirality can be reversed by a spin-polarized current in PMG/Pt heterostructures. Our results enlarge the family of non-reciprocal charge transport systems and pave the way to constructing new chiral spintronics devices.

High-quality PMG films were firstly grown on MgO (001) substrates in the size of 5mm×5mm at 400 °C by co-sputtering Pt and MnGe targets in a high vacuum magnetron sputtering system. Then, the films were left to cool down to room temperature in situ and MgO films were deposited at room temperature to prevent oxidation. Finally, the films were annealed at 500 °C for 2 hours. Figure 1(a) shows the $2\theta$-$\theta$ x-ray diffraction (XRD) pattern measured for 5- and 20-nm-thick PMG films. In addition to the (002) peak of MgO substrate, evident (001) and (002) peaks of PMG can be observed in all the two films. The XRD pattern of 20-nm-thick film in larger angle range has been shown in the inset of Fig. 1(a), and additional (003) peak of PMG has also been found. It should be noted that, based on the intensities of (001) and (002) peaks, the chemical ordering seems to be increased as increasing the thickness from 5 nm to 20 nm. The 360° $\Phi$ scans of the (202) planes of MgO substrate and 20-nm-thick PMG film is shown in supplementary Fig. S1, which indicates the epitaxial growth of the (001)-oriented PMG on the MgO (001) substrate [20]. Fig. 1(b) shows the X-ray reflectivity (XRR) data, which all reveal high crystal qualities, perfect interfaces as well as smooth surfaces. In this analysis, the epitaxial high-quality PMG films are essentially homogeneous and continuous from 5 to 20 nm. According to reciprocal space map of the PMG (204) reflection as shown in Fig. 1(c), the PMG films have shown the relaxation of the epitaxial strain. The in-plane and out-of-plane lattice constants were calculated to be 0.389 nm and 0.391 nm, respectively. In our previous work [21], we have fabricated high quality PtMnGa film on MgO (001) substrate using the same method. Referring to all the XRD results, the PMG film seems to have a similar cubic lattice structure with PtMnGa film. However, the PMG film in this work has a higher chemical ordering, since as compared with PtMnGa film, we have found (001), (002) and (003) peaks of PMG. By the way, the PMG film could also have a random distribution of all three atoms and we cannot confirm the lattice point of each atom. However, it should be noted that the PMG film also does not belong to the full Heusler alloy, since the lattice we have calculated is surprisingly small, which is physically unreasonable as discussed in our previous work [21].

We have carried out the magnetic curves measurements using superconducting quantum interference device (SQUID). It should be noted that we have subtracted the diamagnetic signals of MgO substrates. Firstly, the magnetic curves of the MgO (001) substrate in the size of 5mm×5mm were carried out by SQUID. Then, we have measured the magnetic curves of the PMG/MgO (001) films in the same size. Finally, we have subtracted the contribution of MgO substrate from the total magnetic signals. Fig. 1(d) shows the magnetic curves of the 10-nm-thick PMG film at 300 K for both in-plane and out-of-plane cases. The magnetic-field-induced magnetization has been found for in-plane case, while the magnetization shows a negative linear dependence on magnetic field for out-of-plane case, indicating the diamagnetism signals. The results indicate that the PMG film does not strictly belong to common ferromagnet, ferrimagnet or antiferromagnet. More magnetic characteristics at low temperatures have been shown in supplementary Fig. S2 [20].

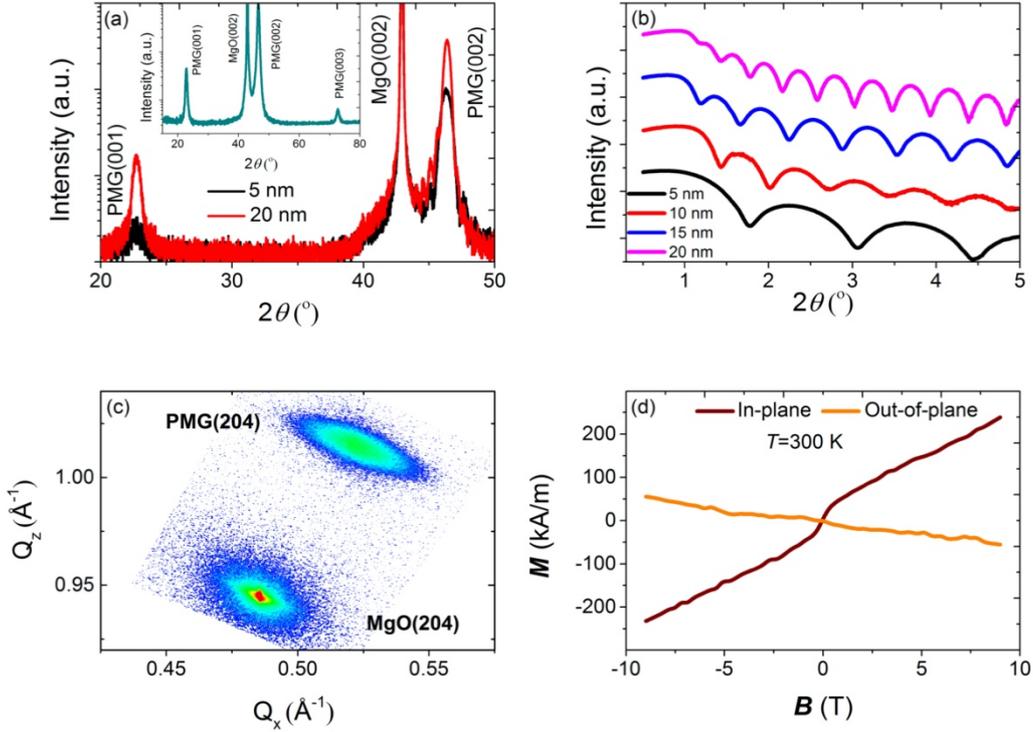

**Figure 1.** 2$\theta$-$\theta$ XRD (a) and XRR (b) measured for MgO (001) (sub.)/PMG/MgO (2 nm) films. The inset of (a) is the XRD pattern of a 20-nm-thick film in larger angle range. (c) High-resolution XRD reciprocal space map of MgO (001) (sub.)/PMG (10 nm)/MgO (2 nm) films. (d) In-plane and out-of-plane magnetic curves of MgO (001) (sub.)/PMG (10 nm)/MgO (2 nm) films at 300 K.

Using XMCD and XMLD at the Mn $L_{2,3}$ edges, we have further examined the magnetic properties of PMG film. The absorption spectroscopy were recorded using total electron yield method from 100 K to 350 K, which directly detected sample electron current while scanning the photon energy. From the x-ray absorption spectroscopy (XAS) shown in Fig. 2 (a), one can find that the distinctly split structure of the $L_2$ edge and the two weak shoulders in the high-energy tail of $L_3$ edge resemble the features in MnPt film [22]. The absence of XMCD signals at both 100 K and 300 K indicates the zero net magnetization. The XMLD was obtained by keeping the direction of the electric vector ***E*** of the incident linearly polarized light fixed in space and rotating the sample as illustrated schematically in Fig. 2(b). The absorption edges recorded for polar rotation ($\beta$=0º and $\beta$=70º) are shown in Fig. 2(c). The features of the spectroscopy can be identified with multiplet structures in a high-spin state with predominant Mn-$3d^5$ configuration [22]. The presence of the XMLD signal which is the difference between the XAS of $\beta$=0° and $\beta$=70° means that the Mn sublattice magnetization is antiparallel in PMG film. The dichroism is in the order of 1% at both 100 K and 300 K, which is similar with MnPt alloys [22]. Referring to the different amplitude variation at $L_3$ edges at 100 K and 300 K, one can exclude the crystal-field effect which can also contributes to the XMLD but is independent on temperature. In contrast with some magnetic oxides [23], the influence of the tetragonal distortion in magnetic alloys film PMG is weak and the structure of PMG should not be conspicuously modified as decreasing temperature. Furthermore, the contribution of crystal-field effects is larger for lower symmetry case [23, 24]. Although random distribution of the three atoms cannot be avoided, the PMG should not have low symmetry. Therefore, we have concluded that the XMLD signals are mainly determined by the relative orientation of magnetic moments and polarization vector of the light. By the way, the very small but nonzero XMLD signal at 350 K indicates that the magnetic transition temperature should be larger than at least 300 K (see supplementary Fig. S3) [20].

The intensity ratio between the $L_3$ well defined double peak structures (labeled as A and B) turns out to be well suited to detect the spectroscopy changes. The intensity ratios shown in Fig. 2(d) reveal that the measurement has polar rotation dependence in the absence of magnetic field. In contrast, we have not found pronounced difference between the XAS spectroscopy for azimuthal rotation ($\phi$=0° and $\phi$=90°) (see supplementary note 1) [24]. A large part of Mn magnetic moments seem to be perpendicular to the surface at zero magnetic field, but there are also in-plane components due to large error bars [25]. Combined with the magnetic curves, The XMCD and XMLD data further indicate that the PMG film does not strictly belong to common ferromagnet, ferrimagnet or antiferromagnet, but its special magnetic configurations could promote the non-reciprocal response as shown in the following discussion.

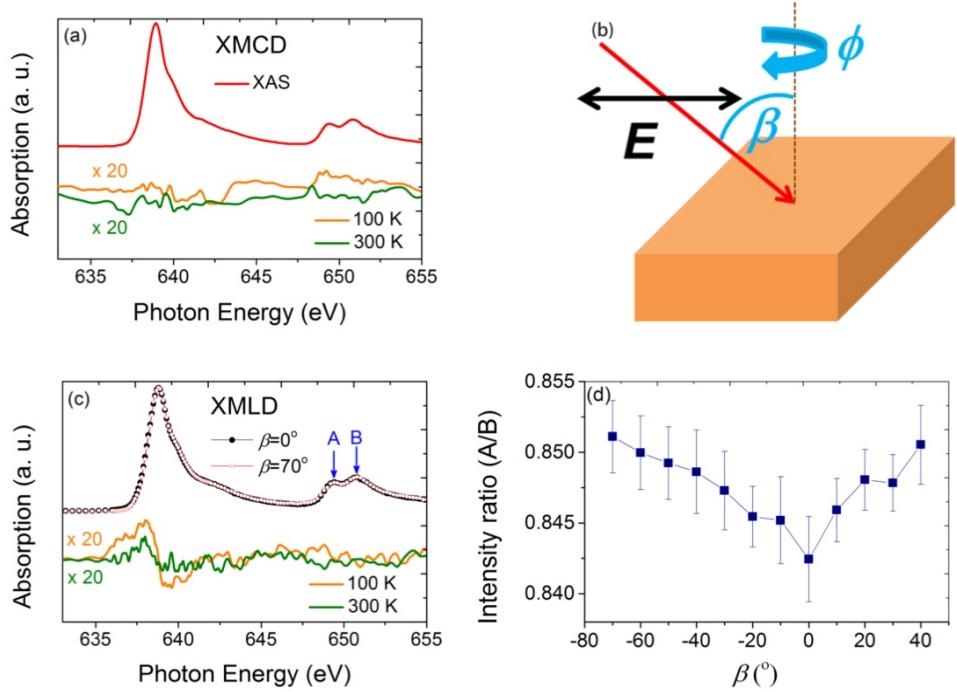

**Figure 2.** (a) Experimental Mn $L_{2,3}$ XAS and XMCD spectroscopies for MgO (001) (sub.)/PMG (10 nm)/MgO (2 nm) films measured at 100 K and 300 K. The spectroscopy is normalized to the $L_3$ peak maximum. (b) Experimental geometry for XMLD measurements and definition of the coordinate system. (c) Mn $L_{2,3}$ XAS spectroscopy for MgO (001) (sub.)/PMG (10 nm)/MgO (2 nm) films as a function of the angle $\beta$ between $\mathbf{E}$ and the surface normal at 300 K. The XMLD spectroscopy at 100 K and 300 K are plotted at the bottom. (d) XMLD spectroscopy for the polar rotations at 100 K. The intensity ratio of the Mn $L_2$ double peak structures is used to measure the intensity of the effect.

Photolithography and Ar ion milling were used to pattern Hall bars, and a lift-off process was used to form the Pt contact electrodes. A scanning electron microscope (SEM) image of a patterned Hall bar and the schematic of the measurement setup along with the definition of the coordinate system used in this study are shown in Fig. 3(a). It consists of three long strips and allows for investigating the non-reciprocal charge transport through applying current along different crystal directions. The size of each strip is 10μm×100μm. The AC, DC and pulsed current transport measurements were carried out in a Quantum Design Physical Property Measurement System (PPMS). The DC current transport has been discussed in supplementary Fig.

S4 [20]. In this work, we have used the two point geometry for longitudinal resistance measurement, and the longitudinal harmonic resistance results based on four point geometry have been shown in supplementary Fig. S5 [20]. A lock-in technique was employed to record the harmonic resistances with applying an AC current $I = I_0 \sin \omega t$ of frequency $\omega/2\pi = 256$ Hz. The sample was rotated in the x-y ($\alpha$) plane under magnetic field $\boldsymbol{B_x}$ along x direction. According to equation (1), the nonreciprocal resistance can be written as $R(\boldsymbol{I}) - R(-\boldsymbol{I}) \propto R_0 \boldsymbol{B} \cdot \boldsymbol{I} \propto R_0 BI \cos\alpha$. Therefore, the voltage can be expressed as:

$$\begin{aligned} V(I) &= IR \\ &= R_0\left(1+\beta B^2\right)I_0 \sin\omega t + \gamma R_0 I_0^2 B \cos\alpha \sin^2\omega t \\ &= R_0\left(1+\beta B^2\right)I_0 \sin\omega t + \frac{1}{2}\gamma R_0 I_0^2 B \cos\alpha[1+\sin(2\omega t - \pi/2)] \end{aligned} \quad (2)$$

The phases of lock-in amplifiers were set to 0 and $-\pi/2$ for the first and second harmonic signal measurements, respectively [21]. The first harmonic resistance can be defined as $R_\omega = R_0\left(1+\beta B^2\right)I_0/I_0 = R_0\left(1+\beta B^2\right)$, representing the linear current-independent resistance. The nonreciprocal resistance is defined as $\frac{1}{2}\gamma R_0 I_0 B \cos\alpha$.

As an illustrative example, the angle $\alpha$ dependent longitudinal and transverse harmonic resistances in 10-nm-thick PMG at 300 K with applying AC current 5 mA and under $\boldsymbol{B_x} = 9$ T are shown in Fig. 3(b) and (c) respectively. The longitudinal first ($R_\omega$) and transverse first ($R_\omega^H$) harmonic resistances reveal the $\sim(\cos^2\alpha)$ and $\sim(\sin\alpha\cos\alpha)$ features respectively, corresponding to the anisotropic magnetoresistance (AMR) and planar Hall effect of the sample. In general, the AMR originates from effects of spin-orbit coupling on the band structure and from scattering from an extrinsic disorder potential due to off-stoichiometry and inter-site swapping between Pt, Mn and Ge atoms, and more discussion about the AMR has been shown in supplementary Fig. S6. Although a cosine term dominates both the longitudinal ($R_{2\omega}$) and the transverse ($R_{2\omega}^H$) second harmonic resistances as shown in Fig. 3(b) and (c) respectively, the determined physics mechanism should be different. Theoretically, the $R_{2\omega}$ consists of three contributions [26-28], namely the EMCA ($R_{2\omega}^{EMCA}$), the magnetothermal effects due to the temperature gradients induced by Joule heating ($R_{2\omega}^{\nabla T}$), and the SOT induced modulation of the total harmonic resistance ($R_{2\omega}^{SOT}$):

$$R_{2\omega} = R_{2\omega}^{EMCA} + R_{2\omega}^{\nabla T} + R_{2\omega}^{SOT} \quad (3)$$

Due to the symmetric behavior of the harmonic resistance with respect to the *x-y* plane, the out-of-plane oscillations driven by the damping-like SOT do not contribute to $R_{2\omega}$, while only Oersted and field-like effective fields in the form of $\sim(\sin\alpha\cos^2\alpha)$ will contribute to $R_{2\omega}$. The magnetothermal effect occurs due to Joule heating and the corresponding quadratic increase of the film temperature. Both Nernst effect and Seebeck effect will induce longitudinal voltage and transverse voltage proportional to $I^2(\boldsymbol{B}\times\nabla T)$ [26-29], giving rise to angular dependence of $R_{2\omega}$ and $R_{2\omega}^H$, respectively. The second harmonic resistances due to $\nabla T_z$ can be expressed as $R_{2\omega} \propto \nabla T_z \sin\alpha$ or $R_{2\omega}^H \propto \nabla T_z \cos\alpha$, and it should be proportional to the magnetic field. Fig. 3(d) shows the magnetic field dependent $R_{2\omega}$ and $R_{2\omega}^H$ when $\alpha$=0º (*I*//*B*). Under large magnetic field, both of the two signals are proportional to $\boldsymbol{B}_x$. Fig. 3(e) shows a linear $R_{2\omega}$-*I* relationship when $\alpha$=0º and $\boldsymbol{B}_x$=9 T. Referring to the above analyses and the harmonic behaviors shown in Fig. 3(b)-(e), we can therefore conduct the following conclusions: i) there is no SOT in single PMG film; ii) there is magnetothermal effect, but its contribution to $R_{2\omega}$ should be very small; iii) the $R_{2\omega}$ scales linearly with the inner product of current and relatively large magnetic field, indicating the emergence of non-reciprocal charge transport. Furthermore, there is no nonlinear planar Hall effect in PMG as discussed in the work of He *et al.* [29]. The nonlinear planar Hall effect originates from the conversion of a transverse nonlinear spin current to a non-linear charge Hall current by applying an in-plane magnetic field. Such spin-to-charge inter-conversion would not take place in the absence of hexagonal warping and particle-hole asymmetry. Therefore, the nonlinear planar Hall effect should not exist in the PMG film since it does not have the indispensable terms as discussed above. On the other hand, given the magnitudes of the electric and magnetic fields, the transverse second harmonic signal is much smaller than the longitudinal counterpart, which does not suits their calculation results as well. The angle and field dependence of second harmonic signals in Fig. 3(c) and

(d) should just stem from the thermal effect such as the Nernst effect. The possible Nernst effect should give rise to the same nonlinear resistivity in the transverse and longitudinal directions. Therefore, even though there is Nernst effect, it is not the dominant mechanism for the measured longitudinal second harmonic signals in the PMG films.

However, there also exist two more notable phenomena. A flatland behavior in $R_{2\omega}$-$\alpha$ curve has been found around $\alpha=90°$ and $\alpha=270°$ as shown in Fig. 3(b). In our previous work, we have also found a similar behavior in Pt/PtMnGa bilayers, where the flatlands are observed to broaden for smaller current [21]. On the contrary, in PMG single film, we have found that the shape of these flatlands is independent on both current and magnetic field as shown in supplementary Fig. S7 [20]. Ignoring the possible quantum transport and semi-classical transport models as discussed in our previous work, here we give a straightforward explanation for this behavior that it may be associated with the motion and pinning of domain walls during rotating the sample [30-32]. When the orientation of current and magnetic field is nearly perpendicular, the trivial electron scattering from domain walls becomes more dominated, annihilating the non-reciprocal charge transport. As shown in supplementary Fig. S8, although a cosine term dominates the $R_{2\omega}$-$\alpha$ curve, the signals around $\alpha=90°$ and $\alpha=270°$ also depend on the device shapes, which could consist of various size and quantity of domain walls [20]. The flattening of the $R_{2\omega}$ is thus taken as a sign of domain walls motion and pinning characteristics. In the latter discussion about the magnetic structures, we will also give another possible explanation for this flatland behavior. The other notable phenomenon is the absence of $R_{2\omega}$ at around zero magnetic fields as shown in Fig. 3(d), which also features a flatland behavior. Here, the domain walls model may also play an important role, but we want to emphasize the role of magnetic field which could induce the long range chiral magnetic order. Fig. 3(f) shows the temperature dependent EMCA coefficient $\gamma$ with applying AC current of 5 mA, and the longitudinal resistance ($R_L$) with applying DC current of 0.5 mA. An increase of the $R_L$ with the increased temperature similar to normal metallic materials has been observed in PMG. The estimated value of $\gamma$ is $\sim$ $-10^{-2}$ $A^{-1}T^{-1}$ in the temperature range from 200 K to 400 K. Notably, the absolute value of $\gamma$ is much larger than those reported so far at room temperature [6-13, 33-35]. The smaller absolute value of $\gamma$ at low temperature is ascribed to the glassy ferromagnetic state in PMG as shown in supplementary notes 2 and 4, which could suppress the chirality dependent charge scattering [20].

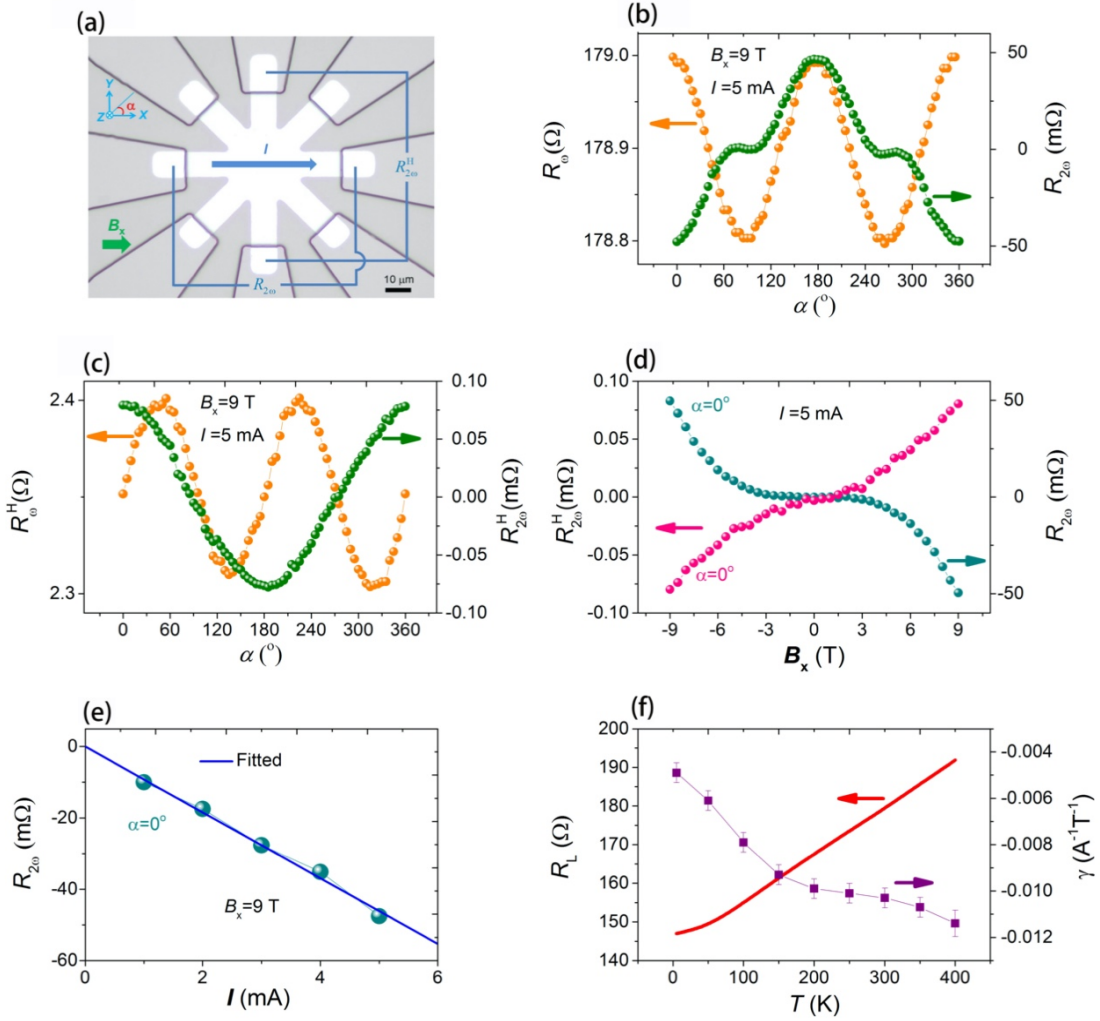

**Figure 3.** (a) A scanning electron microscope image of the device, as well as schematic for longitudinal ($R_{2\omega}$) and transverse ($R_{2\omega}^H$) second harmonic resistance measurements under in-plane magnetic field $B_x$ with rotating the sample in x-y ($\alpha$) plane. Angle ($\alpha$) dependence of the longitudinal (b) and transverse (c) harmonic resistances with applied AC current 5 mA in MgO (001) (sub.)/PMG (10 nm)/MgO (2 nm) films under $B_x$=9 T at 300 K. (d) Magnetic field dependence of longitudinal $R_{2\omega}$

and transverse $R_{2\omega}^{H}$ second harmonic resistances with applied AC current 5 mA in MgO (001) (sub.)/PMG (10 nm)/MgO (2 nm) films at 300 K. (e) AC current amplitude dependence of $R_{2\omega}$ under $\bm{B_x}$=9 T at 300 K when $\alpha$=0°. Solid lines are fits to the experimental data. (f) Temperature dependent longitudinal resistance $R_L$ and EMCA coefficient $\gamma$ in MgO (001) (sub.)/PMG (10 nm)/MgO (2 nm) films.

To confirm the non-reciprocal charge transport but eliminate the thermal effect contributions as much as possible, we have carried out the pulsed current resistance measurement at 300 K, which is schematically shown in Fig. 4(a). After applying a pulsed current along positive *x* direction with the width 50 μs, the longitudinal resistance $R_P$ is measured after a 16 μs delay under an external magnetic field $\bm{B_x}$ along positive *x* direction. The next same measurement was performed with a delay of 10 s for thermal relaxation. After five successive repetitions, the pulse current direction was switched to negative *x* direction, and the longitudinal resistance $R_N$ is measured after a 16 μs delay under the same external magnetic field $\bm{B_x}$ along positive *x* direction. Then the measurement is also delayed for 10 s and another five successive repetitions were conducted. These ten pulses can be regarded as one measurement period. Fig. 4(b) shows the data of $\Delta R/R_L = \left(R_{P(N)} - R_P\right)/R_L$ for five measurement periods with fixed magnetic field $\bm{B_x}$=9 T but different pulsed current amplitude, and Fig. 4(c) shows the data with fixed pulsed current amplitude but different magnetic field. Actually, both $R_P$ and $R_N$ are larger than $R_L$ with applying DC current of 0.5 mA at 300 K due to thermal effect. Therefore, the $R_L$ in above equation is the longitudinal resistance with applying DC current of 0.5 mA at real temperature of the sample. We have compared the values of $R_N$ and $R_P$ with the $R_L$-$T$ curves as shown in Fig. 3(f), and roughly estimated the temperature rise when $R_L$ is approximate to the average value of $R_P$ and $R_N$. For example, the $R_L$ under DC current 0.5 mA at room temperature is 179.38 Ω, while $R_N$=181.97 Ω and $R_P$=181.25 Ω under pulsed current 20 mA at room temperature. Then, according to the $R_L$-$T$ curves, the temperature is about 312.3 K if $R_L$ is ~181.61 Ω. Here, we want to note the role of thermal effect, so we have not directly used the equation $(R_N-R_P)/((R_N+R_P)/2)$ to denote the nonreciprocal response. The data shown in Figs. 4(b) and (c) can be straightforwardly explained by non-reciprocal charge transport, since the following relation is fulfilled:

$$\Delta R/R_L = (R_N - R_P)/R_L \approx -2\gamma IB.$$

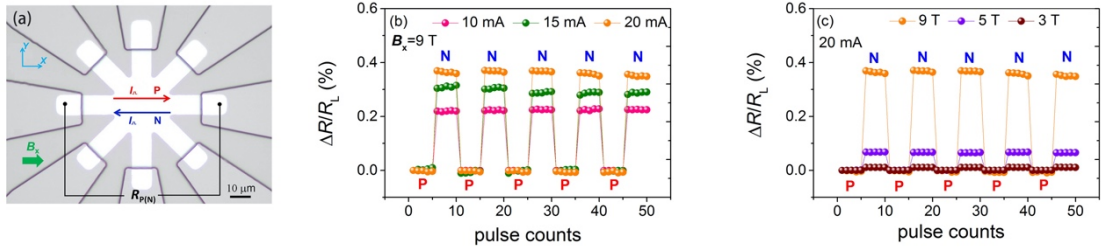

**Figure 4.** (a) Schematic for pulsed current resistance measurements under in-plane magnetic field $B_x$. (b) $\Delta R/R_L = (R_{P(N)} - R_P)/R_L$ for five measurement periods (50 pulse counts) with fixed magnetic field $B_x$=9 T but different pulsed current amplitude. (c) $\Delta R/R_L = (R_{P(N)} - R_P)/R_L$ for five measurement periods (50 pulse counts) with fixed pulsed current amplitude 20 mA (~$10^7$ A/cm$^2$) with varying applied magnetic field $B_x$.

Tuning of the magnetic orders can be realized in multicomponent systems with competing types of interactions such as magnetocrystalline anisotropy and dipole-dipole and Dzyaloshinskii-Moriya interactions (DMI) [36-39]. In the presence of magnetic field, the strong spin-orbit coupling due to the emergence of Pt atoms may dramatically modulate the interaction competitions and bring out a complex magnetic order in PMG such as non-collinear spin configurations, chiral conical magnetic structures and skyrmion, or the coexistence of above [40-45]. Many *B*20 compounds display a helical ground state due to the competition of the DMI with the Heisenberg exchange [43, 44]. In the presence of a uniform field, the helical order unpins and skyrmion lattice is realized, which can approximately be visualized by a simple superposition of three helical states. The non-reciprocal charge transport was recently investigated in a cubic chiral helimagnet MnSi which contains multiple helical axes [8]. It was reported that the EMCA is enhanced at large field above the critical temperature due to chiral spin fluctuations but is significantly suppressed when the chiral conical phase and chiral magnetic skyrmion are predominantly formed. Until now, we have not found evident skyrmion in PMG by means of Lorenz microscopy under magnetic field from 0 T to 3 T. Therefore, the skyrmion lattices should not account for the non-reciprocal charge transport in PMG. A helical or spiral arrangement is given if the magnetic moments are aligned parallel in a plane but the direction varies from plane to plane in such a way that the vector of the magnetic moment moves on a circle or a cone, respectively. Recently, the non-reciprocal charge transport has been investigated in a prototype monoaxial chiral helimagnet $CrNb_3S_6$, which has a homochiral crystalline structure [11]. The EMCA revealed drastic changes when paramagnetic phase, forced ferromagnetic phase and chiral magnetic order of $CrNb_3S_6$ were modulated through varying magnetic field and temperature. Here, on the contrary, the PMG films do not have a homochiral crystalline structure and there is no crystalline contribution to EMCA as compared with $CrNb_3S_6$. Through XMLD and XMCD as discussed above, we can only confirm the orientations of Mn magnetic moments of PMG, but cannot definitely verify whether there are helical or conical magnetic structures under large in-plane magnetic field. Theoretically, it could be impossible to inherently develop these kinds of magnetic structures since there is a close correlation between the structural chirality and the chiral magnetic orders. However, according to the magnetic characteristics and transport measurements, the PMG should have non-zero vector spin chirality $S_i \times S_j$ due to non-collinear spin configurations under large in-plane magnetic field, which was theoretically reported to scatter electrons asymmetrically, resulting in non-reciprocal transport phenomena [46]. The combination of the non-zero vector spin chirality and the non-zero magnetization in PMG under large in-plane magnetic field determines the nonreciprocity. This mechanism could also explain the flatland behavior in $R_{2\omega}$-$\alpha$ curve around $\alpha$=90º and $\alpha$=270º as shown in Fig. 3(c). Considering the asymmetric scattering is proportional to the vector spin chirality, the achiral spin dependent scattering will become more evident when the current gradually orient to the perpendicular direction of $S_i \times S_j$. Furthermore, a non-zero scalar spin chirality $S_i \cdot (S_j \times S_k)$ may also exist under out-of-plane magnetic field, which is closely related to the AHE at room temperature in PMG as discussed in supplementary note 4 [20]. More theoretical and experimental considerations on the microscopic origin of the chirality in PMG especially under large magnetic field are eagerly required.

Furthermore, we have also demonstrated that the vector spin chirality can be controlled by a spin-polarized current of Pt through spin Hall effect in PMG/Pt bilayers. The preparation method is the same with our previous work [21]. According to the TEM results of the PtMnGa/Pt bilayers, we think the interdiffusion in this case is also weak [21]. Fig. 5 shows the non-reciprocal charge transport in PMG(5 nm)/Pt(5 nm) bilayers under magnetic field of 9 T at 300 K. As shown in Fig. 5(a), the chirality has been gradually reversed as increasing the applied current, featuring the sign change of EMCA coefficient $\gamma$ characteristics. It seems that, under large applied current, two types of chirality coexist in the film systems since the chirality reversal happens only when the current is almost parallel or antiparallel with magnetic field. In Fig. 5(b), we have compared the field dependent $R_{2\omega}$ between 5 mA and 9 mA when $\alpha=0°$. It is found that only one type of chirality exist for the case of 5 mA, while a chirality reversal has been found for 9 mA ($j_c=4.5\times10^6$ A/cm$^2$). However, it also reveals that the chirality reversal needs not only large current but also large magnetic field. By the way, we have not gotten the results with much higher applied current since such a high-density electric current has outranged our harmonic systems and damaged the samples. The illustration of the AC current control of chirality in PMG/Pt bilayers has been discussed in supplementary note 6 [20]. The chirality reversal using a spin-polarized current at room temperature should enrich the area of solid-state physics and microelectronics by the range of unique characteristics of non-collinear magnetic structures. However, there are still a few issues to be resolved in future to realize such application, and one is the use of high magnetic fields. In this work, we have applied a magnetic field to control and to detect the chirality, but it cannot be used in practical electronic devices. Now, we are carrying out more detailed studies on the chirality reversal based on PMG, which may be present in our future research report.

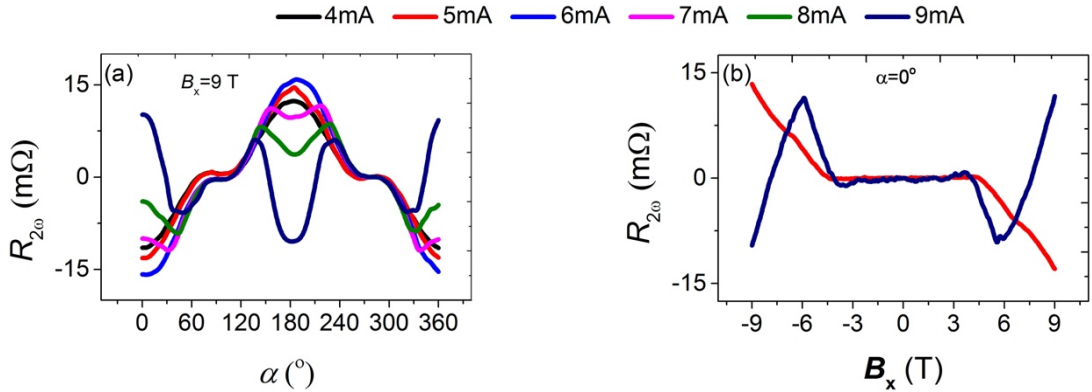

**Figure 5.** Angle (a) and in-plane magnetic field (b) dependence of the longitudinal second harmonic resistances $R_{2\omega}$ with varying applied AC current in MgO (001) (sub.)/PMG (5 nm)/Pt (5 nm) films at 300 K.

In conclusion, we have fabricated high quality magnetic alloys film PMG and confirmed the large non-reciprocal charge transport up to room temperature through AC harmonic and pulsed current transport measurements. We ascribe this chiral dependent transport characteristic to the non-zero vector spin chirality in the presence of magnetic field, which can also be reversed through a spin-polarized current. We envision our findings paving a new route to the development of new functional electronic devices.

**Acknowledgements:** We thank D. Z. Hou and Q. Li for the useful discussions. This work was partially supported by the National Natural Science Foundation of China (Grants No. 51971027, No. 51731003, No. 51971023, No. 51927802, No. 51971024, No. 52061135205), Beijing Natural Science Foundation Key Program (Grant No. Z190007), and the Fundamental Research Funds for the Central Universities (Grants No. FRF-TP-19-001A3, FRF-MP-19-004, FRF-BD-20-06A, FRF-BD-19-010A).